\documentclass[traditabstract]{aa}
\usepackage{txfonts}
\usepackage{graphicx}
\usepackage{epsfig}

\newcommand{\be}{\begin{equation}}
\newcommand{\ee}{\end{equation}}

\authorrunning{Lachowicz \& Done}
\titlerunning{QPOs under wavelet microscope}

\begin{document}

\title{Quasi-periodic oscillations under wavelet microscope: the application of
Matching Pursuit algorithm}

\author{Pawe{\l}~Lachowicz\inst{1,2} \and Chris~Done\inst{3}}

\offprints{P.~Lachowicz\\ \email{pawel@ieee.org}}

\institute{
Nicolaus Copernicus Astronomical Centre, Polish Academy of Sciences,
ul.~Bartycka~18, 00-716 Warszawa, Poland
\and
Centre for Wavelets, Approximation and Information Processing,
Temasek Laboratories at National University of Singapore,\\ 5A Engineering Drive
1, \#09-02 Singapore 117411
\and
Department of Physics, University of Durham, Science Laboratories, South Road,
Durham, DH1 3LE, England
}

\date{Accepted: Mar 05, 2010}

\abstract{We zoom in on the internal structure of the low-frequency
quasi-periodic oscillation (LF QPO) often observed in black
hole binary systems to investigate the physical nature of the lack of
coherence in this feature. We show the limitations of standard Fourier
power spectral analysis for following the evolution of the QPO with
time and instead use wavelet analysis and a new time-frequency
technique -- Matching Pursuit algorithm -- to maximise the resolution
with which we can follow the QPO behaviour. We use the LF QPO seen in
a very high state of XTE~J1550$-$564 to illustrate these techniques and
show that the best description of the QPO is that it is composed of
multiple independent oscillations with a distribution of lifetimes but
with constant frequency over this duration. This rules out models
where there is continual frequency modulation, such as multiple blobs
spiralling inwards. Instead it favours models where the QPO is excited
by random turbulence in the flow.}


\maketitle

\section{Introduction}

One of the most outstanding features of the stellar mass black-hole
binaries (BHB) is their rapid X-ray time variability 
(e.g. van der Klis 1989). This variability
consists of aperiodic continuum noise spanning a broad range in
frequencies, with quasi-periodic oscillations (QPOs) superimposed.
The `low frequency QPO' (hereafter LFQPO) is the one most often
seen, and this is at a characteristic frequency which {\em moves} from
0.1--10~Hz in a way which is correlated with the energy spectrum of the
BHB (see e.g. the reviews by van der Klis 2006; McClintock \&
Remillard 2006; Done et al. 2007).

The true nature of this LFQPO (or any of the other QPO's) is still not
well understood, and there are multiple suggestions in the literature.
The most fundamental frequency is Keplarian, but this is very high, at
$\sim 200$~Hz at the last stable orbit for a $10 M_\odot$ Schwarzchild
black hole (Sunayev 1973). Instead, the QPO could be linked to the
buildup and decay of `shots' on the longer timescales for magnetic
reconnection events (Miyamoto \& Kitamoto 1989; Negoro et al. 1994),
though to form a QPO probably requires correlation between the shots
via an energy reservoir (Vikhlinin et al. 1994). Alternatively, the frequency
could be set by modes excited in
the accretion disc (Nowak \& Wagoner 1995; Nowak et al. 1997), or by
one-armed spiral waves in the disc (Kato 1989) or Lense-Thirring
precession (Ipser 1996; Stella \& Viertri 1998: Psaltis \& Norman
2000; Ingram et al. 2009).

Clearly, observational constraints are required in order to
distinguish between some of these very different physical models.
Firstly, the energy dependence of the QPO clearly shows that the
oscillation does not involve the optically thick accretion disc
emission but is instead a feature of the higher energy Comptonised
tail (e.g. {\.Z}ycki et al. 2007). Therefore all models using a modulation of
the
{\em thin disc} are ruled out.  Instead, the QPO must be connected to the hard
X-ray emitting coronal material, though the `shot noise' models are ruled
out by the observed {\it r.m.s.}--flux correlation as this cannot be
formed by independent events (Uttley et al. 2005).

More constraints come from the detailed shape of the QPO signal. The
most popular method for studying X-ray variability is the Fourier
power spectrum density (PSD; see Vaughan et al. 2003 and references
therein). This is the square of the amplitude of variability at a
given frequency, as a function of frequency, namely $P(f)$. Power
spectra from BHB can be very roughly described as band limited noise,
with a `flat top' in $\nu P(\nu )$ (equal variability power per decade
in frequency) i.e. $P(\nu)\propto\nu^{-1}$. This extends between a low
and high frequency break, $\nu_{b}$, below which the PDS is
$P(\nu)\propto \nu^0$, and $\nu_{h}$, above which the spectrum
steepens to $P(\nu)\propto \nu^{-2}$. However, this broadband noise is
better described by a sum of 4--5 Lorentzian components (Belloni \&
Hasinger 1990; Psaltis et al. 1999; Nowak 2000; Belloni, Psaltis \& van der Klis
2002). This has the advantage that LF
QPO is also well modelled by a Lorentzian, so both QPO and underlying
noise can be fitted with the same functions. More importantly,
Lorentzians also suggest a {\em physical} interpretation of the power
spectral components as these are the natural outcome of a damped,
driven harmonic oscillator. Each Lorentzian-like structure in a PSD may
then be related to a process in the accretion flow, hence X-ray
emission $x(t) \propto \sin(2\pi f_0 t)\exp(-t/\tau)$ where $\tau$
denotes a damping time-scale (Misra \& Zdziarski 2008; Ingram et al.
2009).

A Lorentzian has $P(f)= A \Delta f/[(f-f_0)^2 +(\Delta f/2)^2]$,
defined by a centroid frequency, $f_0$, a full width at half maximum
(FWHM) of $\Delta f$ and amplitude.  A ratio of the QPO frequency to
Lorentzian FWHM is known as a quality factor, $Q=f_0/\Delta f$. The
distinguishing feature of the QPO as opposed to the noise components
is that its quality factor can be large. All these paramters are
correlated in the LF QPO, with the amplitude and quality factor
increasing along with the increase in centroid frequency (e.g. Nowak
et al. 1999; Nowak 2000; Pottschmidt et al. 2003).

It is obviously important to know what sets the (lack of) coherence of
the QPO. Are QPOs composed of a set of longer-lasting continuous
modulations or are they rather fragmented in time with frequencies
concentrated around the peak QPO frequency? Is there any correlation
between observed oscillations or they are preferably excited in a
random manner?  To answer these questions we need to resolve and
understand the internal structure of detected QPOs i.e. to study the
behaviour of the QPO on timescales which are {\em short} compared to its
observed broadening. A continuous modulation would then show up as a
continuous drift in QPO frequency with time, while a series of short
timescale oscillations will show up as disjoint sections where the
QPO is on and off.

One way to do this is using dynamical power spectra (spectrograms; hereafter
also referred to as STFT for simplicity). While the actual techniques for this
can be quite sophisticated, (e.g. Cohen 1995; Flandrin 1999), in essence, an
observation of length $T$ sampled every $\Delta t$ (giving a single
power spectrum spanning frequencies $1/T$ to $1/(2\Delta t)$ in steps
of $1/T$) is spilt into $N$ segments of length $T/N$. This gives $N$
independent power spectra spanning frequencies $N/T$ to $1/(2\Delta
t)$ but crucially, the resolution is now lower, at $N/T$. While
this is useful in tracing the evolution of the QPO (e.g. Wilms et
al. 2001; Barret et al. 2005), it introduces `instrumental' frequency
broadening from the windowing of the data which prevents us following
the detailed behaviour of the QPO on the required timescales. 

The real problem with such Fourier analysis techniques is that they
decompose the lightcurve onto a basis set of sinusoid
functions. These have frequency $f$, with resolution
$\Delta f=N/T$, but exist everywhere in time across the duration of
the observation $t_{dur}=T/N$. In the same way that the Heisenburg
uncertainty principle sets a limit to the measurement of momentum and
location of a particle, namely $\Delta p_x \Delta x \geq h/2\pi$ where
$h$ is a Planck constant, there is a Heisenberg-Gabor uncertainty
principle setting the limiting frequency resolution for time-series
analysis. This states that we cannot determine both the frequency and
time location of a power spectral feature with infinite
accuracy (e.g. Flandrin 1999). 

Instead, if the QPO is really a short-lived signal, we will gain in
resolution and get closer to the theoretical limit by using a set of
basis functions which match the underlying physical shape of the
QPO. This is the idea behind {\em wavelet analysis}. For example, one
particular basis function shape is the Morlet wavelet, which is a
sinusoid of frequency $f$, modulated in amplitude by a gaussian
envelope such that it lasts only for a duration $t_{dur}$. The product
$f \times t_{dur}$ is set at a {\em constant}, fixing the number of
cycles seen in the basis function. The lightcurve is then decomposed
on these basis functions, calculated over a set of frequencies so that
the basis function shape is maintained (i.e. that the oscillation
consists of 4 cycles, so low frequencies have longer $t_{dur}$ than
high frequencies). The resolution adjusts with the frequency,
making this a more sensitive technique to follow short duration
signals (e.g.  Farge 1992; Addison 2005).

Quite quickly, wavelet transforms gained huge popularity within the
astronomical community (e.g. Szatmary, Vinko \& Gal 1994; Frick et
al. 1997; Aschwanden et al. 1998; Barreiro \& Hobson 2001; Freeman et
al. 2002; Irastorza et al. 2003) whereas
its application in X-ray astronomy was not so rapid. Scargle et
al. (1993) used wavelets to examine QPO and very low-frequency noise
in Sco~X-1.  Steiman-Cameron et al. (1997) supplemented Fourier
detection of quasi-periodic oscillations in optical light curves of GX
339--4 by wavelets while Liszka et al. (2000) analysed the ROSAT light curve of
NGC~5548 in short
time-scales. Recent studies of QPOs in X-ray black-holes systems can
be found e.g. in Lachowicz \& Czerny (2005), Espaillat et al. (2008)
and Gupta et al. (2009).

However, the problem is that the basis functions chosen in wavelet
analysis may not be appropriate. For example, we assumed above that
these functions have a fixed shape, lasting for a fixed number of
oscillations at all frequencies. In practice, this may not be the best
description of the QPO. Perhaps the QPO is made from a set of signals
which have a distribution of durations, where $f\times t_{dur}$ is not
constant. We will maximise the resolution with which we can look at
the QPO if and only if we use basis functions which best match its
shape.

To do this we propose the application of the {\em Matching Pursuit}
algorithm (MP).  This is an iterative method for signal
decomposition which aims at retrieving the maximum possible theoretical
resolution by deriving the basis functions from the signal itself. We
specifically use this as an extension of the wavelet technique by
setting the MP basis functions as Gaussian amplitude modulated
sinusoids as before, but allowing the product $f\times t_{dur}$ to be
a free parameter (Gabor atoms).

We apply both wavelet and Matching Pursuit analysis to zoom in on the
detailed structure of the LF QPO. A technical description of these
time-frequency techniques we provide in the Appendix. We 
describe the data used in Section \ref{s:selection}.  This is a {\it
RXTE}/PCA observation of the BHB XTE~J1550--564 which which displays a
strong and narrow QPO at $\sim$4~Hz. Section \ref{s:analysis} shows
how wavelets and MP give more information than the standard dynamical
power spectra. We study wavelet and Matching Pursuit feedback to different QPO
models in Section~\ref{s:mpqpo} whereas in Section~\ref{s:discuss} we discuss
the new physical insights this gives about the QPO mechanism.

\begin{figure}
\includegraphics[angle=0,width=0.49\textwidth]{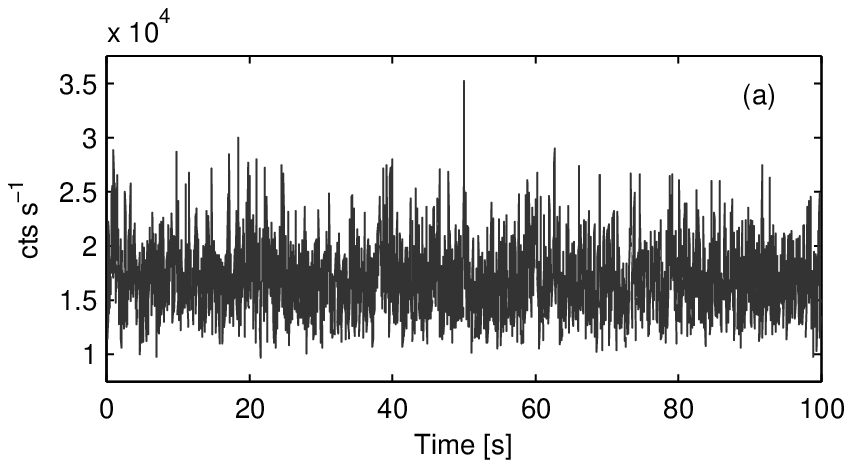}
\includegraphics[angle=0,width=0.48\textwidth]{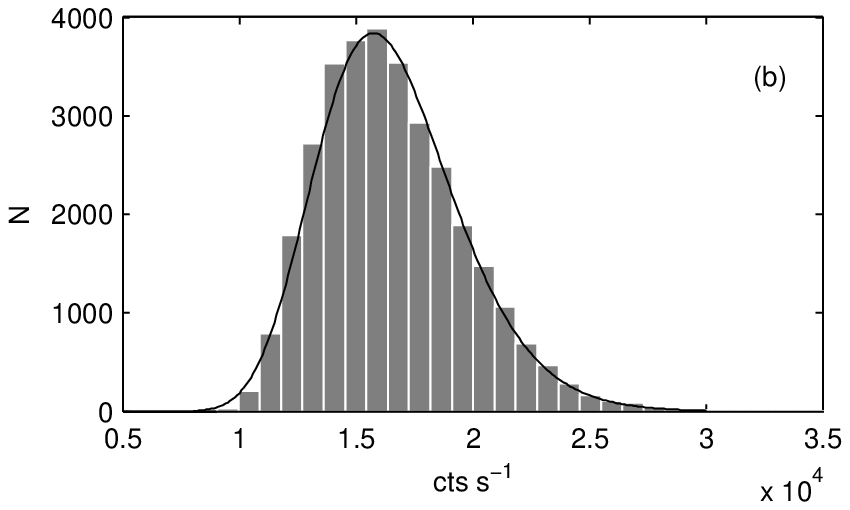}
\caption{(a) An exemplary 100~s data segment extracted from
the light curve of XTE~J1550--564 as observed by {\it RXTE}/PCA on September
29th 1998. (b) The histogram of data in our sample (1000~s) fitted with a
lognormal distribution.}
\label{fig:data}
\end{figure}

\section{Source selection and data reduction}
\label{s:selection}

XTE~J1550--564 is X-ray nova and black-hole binary discovered on September 6,
1998 (Smith et al. 1998) by All Sky Monitor on-board the {\it Rossi X-ray
Timing Explorer}  facility. A possible optical counterpart was identified to be
a low-mass star (Orosz et al. 1998) and the distance to the source was estimated
to be about
5.3 kpc (Orosz et al. 2002). A variable radio source was subsequently found
at the optical position (Campbell-Wilson et al. 1998) and confirmed by Marshall
et al. (1998) based on {\it ASCA} observations. Radio jets with apparent
superluminal velocities were observed after the strong X-ray flare in September
1998 (Hannikainen et al. 2001), placing XTE~J1550--564 among already known
microquasars.

The source 2.5--20 keV spectrum was similar to the spectra of sources that are
dynamically established to be black-holes. XTE~J1550--564 has been observed
in the very high, high/soft and intermediate canonical outburst states of
black-hole X-ray novae (Sobczak et al. 1999a, 1999b; Cui et al. 1999).
The source gained additional support for its black-hole nature due to the
detection of a variety of low-$f$ QPOs (0.08--18~Hz) as well as high-$f$ QPOs
(100--285~Hz) during some of the PCA/{\it RXTE} pointings (Bradt et al. 1993;
Cui et al. 1999; Remillard et al. 1999; Wijnands et al. 1999; Homan et al.
2001).

To probe the object quasi-periodic variability in time-scales of
seconds we use a pointed observation of the Proportional Counter Array
detector of {\it RXTE} taken from the public archive of HEASARC
(http://heasarc.gsfc.nasa.gov). After Zdziarski \& Gierlinski (2004), we
selected the PCA data set of ObsID of 30191-01-15-00 containing source
observation on 29.09.1998 (MJD 51085) when the binary system was in its very
high state (Gierlinski \& Done 2003) and displayed narrow QPO feature of quality
factor $Q\simeq 11$ (Sobczak et al. 2000).

We reduced the data with the LHEASOFT package ver.~6.6.1 applying the standard
data selection: the Earth elevation angle $>10^{\circ}$, pointing offset
$<0^{\circ}.01$, the time since the peak of the last South Atlantic Anomaly SAA
$>30$~min and the electron contamination $<0.1$. The number of Proportional
Counting Units (PCUs) opened during the observation was equal five.

We extracted a light curve in 2.03--13.06 keV energy band (channels
0--30) with a bin size of $\Delta t\!=\!2^{-5}$~s for the first part of
PCA observation (2502~s). For the purposes of this paper, we limited our
data sample only to the first 1000~s dividing it into ten 100~s ($N=3200$
points) segments for which the wavelet and Matching Pursuit analysis was
conducted in detail. An exemplary data segment is presented in
Fig.~\ref{fig:data}a. Our sample returned the mean countrate and r.m.s.
variability equal $1.65\times 10^4$ cts s$^{-1}$ and 18.9\%, respectively. The
data histogram (Fig.~\ref{fig:data}b) can be fitted with a lognormal
distribution, 
\begin{equation}
 f(x;\mu,\sigma) = \frac{1}{x\sigma\sqrt{2\pi}} \exp\left[
	\frac{(-\ln x - \mu)^2}{2\sigma^2} \right] ,
\end{equation}
and the best fit returns the distribution's parameters, the mean and standard
deviation, equal $\mu=9.698\pm 0.003$ and $\sigma=0.184\pm 0.002$, respectively,
where the errors are given at 99\% confidence level (see also Uttley et al.
2005 for a detailed discussion on X-ray light curve distribution).

We computed Fourier power spectra using the {\sc powspec} subroutine in HEASoft
package with Poisson noise level subtraction applied. STFT spectrogram plots
are derived based on the Matlab$^{\mbox{\small \textregistered}}$ Signal
Processing Toolbox ver.~6.9. For the wavelet analysis the Time-Frequency Toolbox
(http://tftb.nongnu.org) was used whereas all results corresponding to the
Matching Pursuit signal decomposition were computed using MP ver.~4 code
provided by Piotr Durka of the Department of Biomedical Physics at Warsaw
University, Poland (http://www.eeg.pl/software).

\section{Time-frequency analysis of 4~Hz QPO structure}
\label{s:analysis}

\begin{figure}
\includegraphics[angle=0,width=0.48\textwidth]{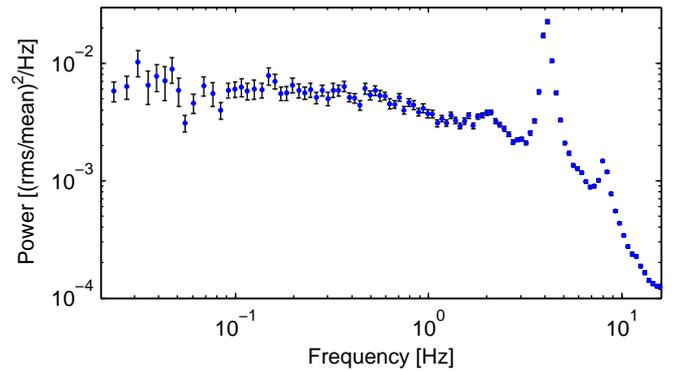}
\caption{Power spectral density plot calculated based on the entire light
curve. Main QPO feature is well detected at the frequency of 4~Hz.}
\label{fig:psd}
\end{figure}

Fig.~\ref{fig:psd} displays the Fourier power spectral density (PSD)
calculated from 78 averaged spectra based on 1024 point data segments. A QPO
peaking at a frequency of 4~Hz is clearly evident together with its harmonic
structure, and r.m.s. variability calculated in a 3.5$-$5 Hz frequency band
equals 12.1\%.

\begin{figure}
\includegraphics[angle=0,width=0.48\textwidth]{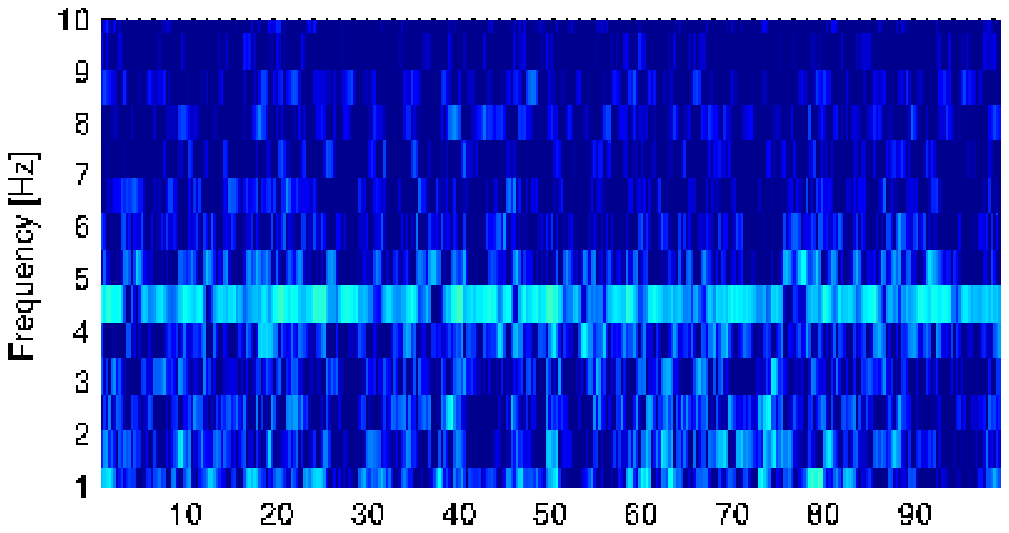}
\includegraphics[angle=0,width=0.49\textwidth]{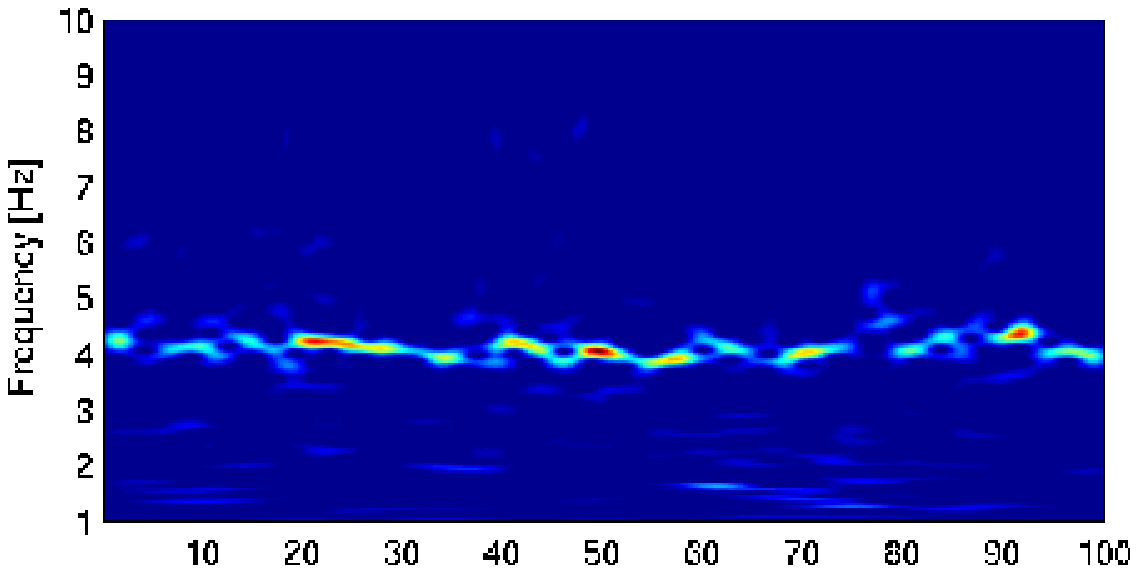}
\includegraphics[angle=0,width=0.49\textwidth]{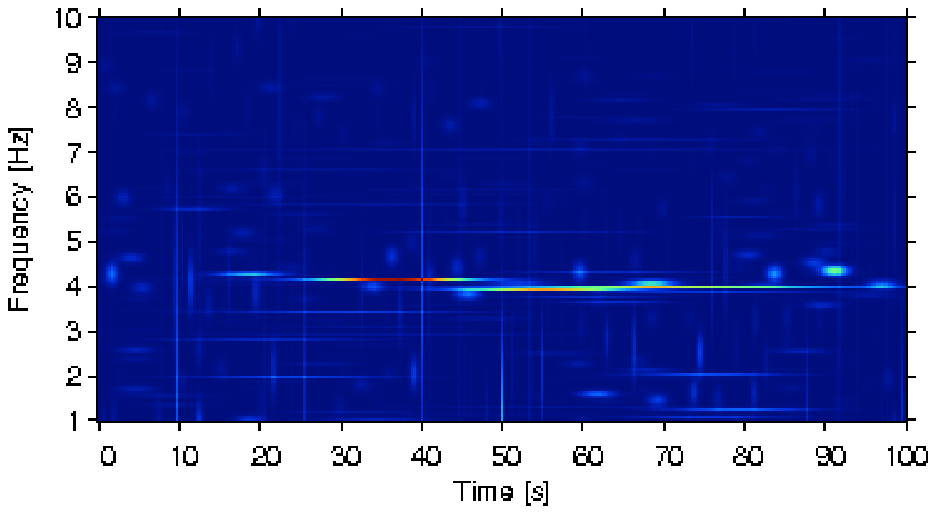}
\caption{Time-frequency analysis performed for the XTE~J1550--564 light
curve presented in Fig.~\ref{fig:psd}a: (upper, a) Spectrogram calculated
for the first 100~s long data segment revealing QPO activity around
a central frequency of 4~Hz (46 point sliding window); (middle, b)
the corresponding wavelet power spectrum derived with a Morlet analysing
function; (bottom, c) Matching Pursuit decomposition (based on $3\times 10^6$
atom dictionary) displayed as energy density $(Ex)(t,f)$ (Eq.~\ref{mpeng})
uncovering the main quasi-periodic components located mainly between
3.7--4.7~Hz. For all figures colour coding is assumed as follows:
increasing values of the STFT/wavelet/MP power (energy density) are denoted by a
gradual brightening of the colour, i.e. from dark blue, yellow to red (colour
values not in scale).}
\label{fig:wamp}
\end{figure}

We use the first 100~s of data (Fig.~\ref{fig:data}a) to compare the three
different time-frequency techniques.  The upper plot in Fig.~\ref{fig:wamp}
shows a standard STFT spectrogram, computed on a 1.44~s long rectangular
(sliding) window\footnote{We select a 1.44~s window size to allow for
a better comparison of the time-frequency QPO structure resolution between the
spectrogram and the wavelet analysis at a frequency of 4~Hz.}. Thus all
oscillations of lifetime shorter than $\sim\!1.44$~s are smeared, and the
frequency resolution is 0.7~Hz, comparable to the FWHM of the QPO. The QPO
appears to be fairly continuous thoughout the dataset, with small excursions
around 4~Hz.

By comparison, the middle panel of Fig.~\ref{fig:wamp} shows the
wavelet power spectrum using the Morlet wavelet basis functions, i.e.
modulated sinusoid confined by a Gaussian envelope, $\psi(t)=\pi^{-1/4}
e^{i(2\pi f_0 t)}e^{-t^2/2}$, with $2\pi f_0=6$ to make each wavelet last
for about 4 cycles irrespective of the timescale $t$ for the oscillation.
At the QPO frequency of 4~Hz (wavelet scale $a\!=\!0.2427$) the
wavelets can resolve signals which last for 1.44~s. Across the QPO width, the
wavelet window changes its duration from 1.63~s at 3.5~Hz to 1.14~s at 5~Hz
causing a continuous change of frequency resolution from 0.61~Hz to 0.88~Hz,
respectively. This gives a better view of the detailed structure of the QPO than
the STFT spectrogram, but still limits our view of the detailed composition of
the QPO signal to half that of the QPO's FWHM. Nontheless, this is already able
to show that the apparently continuous QPO in the STFT result is actually
composed of distinct events. 

\begin{figure}
\includegraphics[angle=0,width=0.475\textwidth]{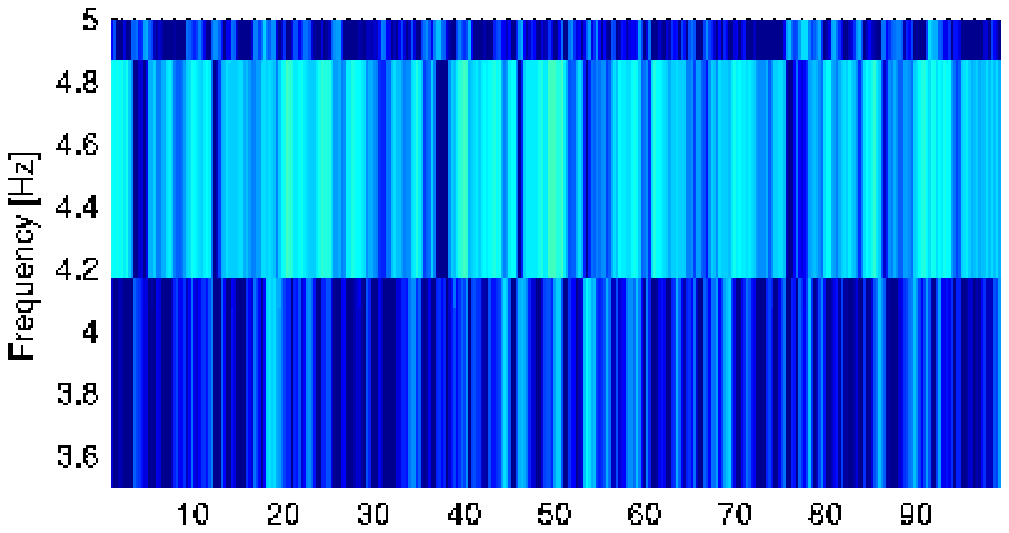}
\includegraphics[angle=0,width=0.475\textwidth]{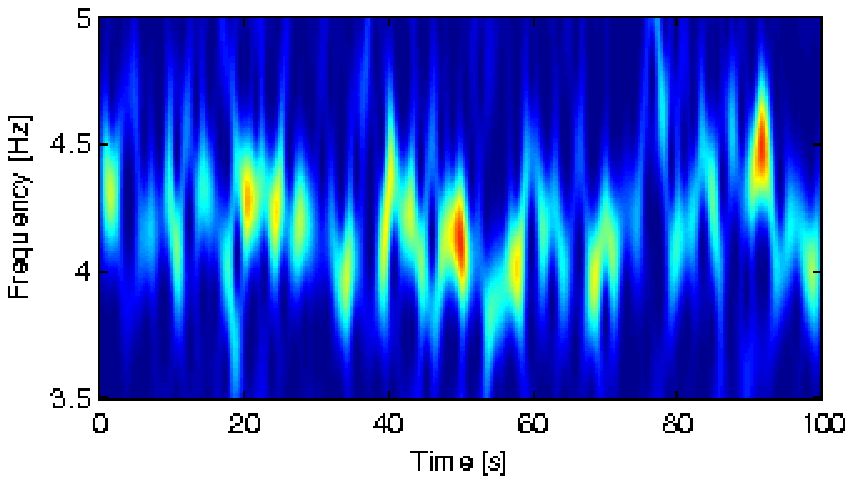}
\includegraphics[angle=0,width=0.475\textwidth]{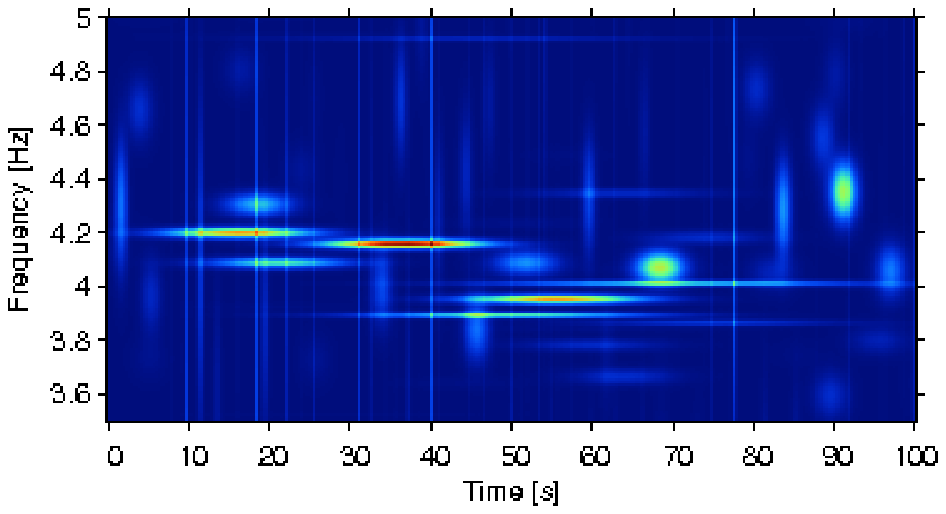}
\caption{As in Fig 2, but zooming in on the 3.5--5~Hz frequency band
  dominated by the QPO.
}
\label{fig:zoom}
\end{figure}

Results from the MP algorithm (Section \ref{ss:mp}) are shown in the
bottom panel of Fig.~\ref{fig:wamp}. This is deconvolved using 500
iterations and displayed as energy density (Eq.~\ref{mpeng}). This is
equivalent to saying that the signal is approximated by 500 separate
Gabor atoms (hereafter the term {\it atom} will be used interchangeably).
Together these account for over 95\% of total signal
power\footnote{Before each MP decomposition we normalise the signal to have
unit energy. Signal energy carried by a single Gabor atom (also
referred to as an {\it atom energy}) is $|\langle R^ix,g_{\gamma,i}\rangle|^2$
and the sum over all atoms (Eq.~\ref{engcon}) tends to unity.}, or speaking
conversely, we can reproduce the entire light curve in 95\% based on 500 fitted
Gabor atoms (figure not shown). The energy density has been displayed with a
linear scale in order to show the time-frequency distribution of the strongest
events.

Fig.~\ref{fig:zoom} shows the detailed structure of the QPO (the
3.5--5~Hz band) derived from each of these techniques. This immediately
shows the progressive increase in time--frequency resolution from STFT
to wavelet to MP analysis.  Since MP algorithm scans the whole stochastic
dictionary for the best matching Gabor functions to the signal, we
gain a new opportunity to represent the QPO structure by a number of
localised periodic functions as defined by Eq.~(\ref{gabor}). The MP
map shows that much of the power of the QPO signal in these data is
made from 6 to 10 periodic oscillations around 4~Hz, with life-times
between 20--60~s. We will return to this result later.

The underlying broadband continuum noise (dominated by the intrinsic
variability rather than purely Poisson statistics, see Fig.~1) is also
present in all these plots. This appears in the STFT as the gradation of
background colour from pale to dark from the bottom (1~Hz) to the top (10~Hz) of
Fig.~\ref{fig:wamp}a. In the wavelets, this is seen as pale, random smudges
around 1$-$3~Hz (Fig.~\ref{fig:wamp}b), whereas in MP they are very short-lived
signals (less than 0.5~s) forming vertical features across the plot. 

\begin{figure}
\includegraphics[angle=0,width=0.475\textwidth]{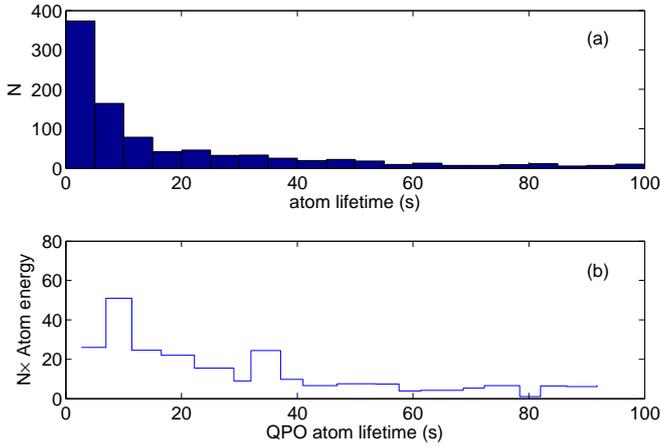}
\caption{The upper panel shows the histogram of atom life-times as calculated by
the MP decomposition within the 3.5$-$5~Hz band. This clearly peaks at
short timescales, but extends continuously up to the longest timescales
sampled by the data. The lower panel shows the power carried by
QPO atoms of each timescale (see text on atom selection). Unlike the atom
lifetimes, this does not peak at zero, but instead shows that most of the QPO
power is carried by oscillations lasting $\sim$5$-$10~s.}
\label{fig:hist_data}
\end{figure}

\section{Detailed QPO structure from Matching Pursuit}
\label{s:mpqpo}

We can quantify the composition of the QPO by showing the distribution
of the life-times of 500 Gabor atoms found in the MP decomposition
accumulated for ten segments of XTE~J1550--564 data in 3.5$-$5~Hz band
(Fig.~\ref{fig:hist_data}a). These show a continuous distribution of lifetimes,
peaked at zero but extending out to the longest timescales sampled by the data.
Fig.~\ref{fig:hist_data}b shows the distribution of power carried by
QPO atoms of each lifetime (sum of atom energy times number of atoms
over all the events in the time bin of that duration). By QPO atoms we define
these events distributed between 3.7$-$4.7~Hz\footnote{The frequency
interval of a width corresponding to the QPO's full width at half maximum.} and
exclude these with a duration below 0.5~s, or with energy which is less than the
events in the 2.5$-$3~Hz
bandpass (reference red-noise level; see Fig.~\ref{fig:psd}) as
both of these will predominantly trace the broadband noise. 

We found that the largest number of atoms (more than half of the total 500 per
data segment) are characterised by a time duration of 0.5$-$3.5~s, 
but that these together carry only half the energy of the strongest 10 atoms,
which have lifetimes of 20$-$60~s (see Fig.~\ref{fig:zoom}c). On the other
hand, Fig.~\ref{fig:hist_data}b reveals that the majority of
QPO power is carried by QPO events with a duration of between 5--10~s.

We now explore to what level the results can distinguish between
different models for the broadening of the QPO signal. 

\begin{figure}
\includegraphics[angle=0,width=0.475\textwidth]{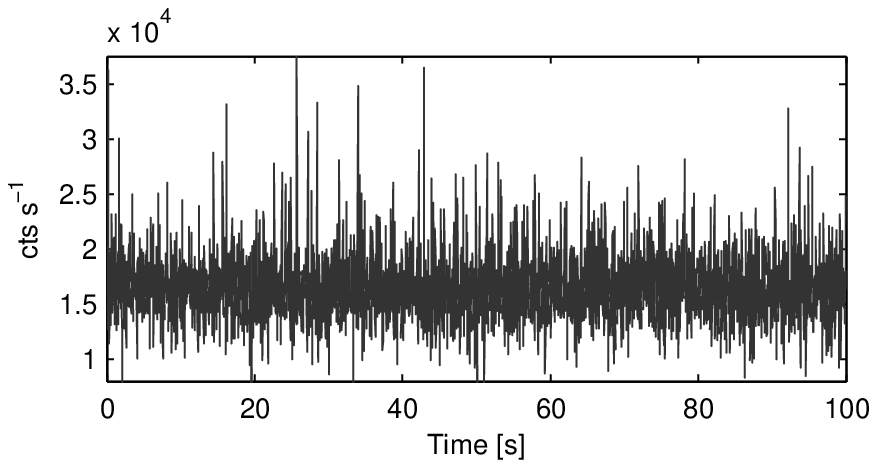}
\includegraphics[angle=0,width=0.475\textwidth]{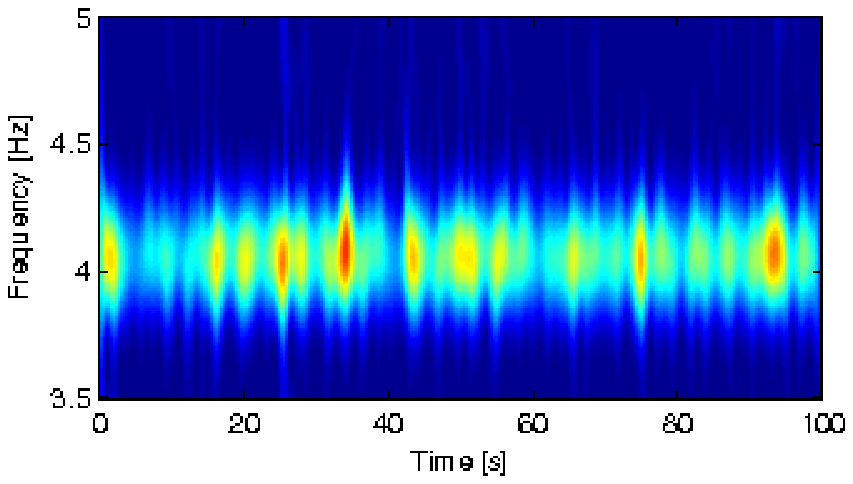}
\includegraphics[angle=0,width=0.470\textwidth]{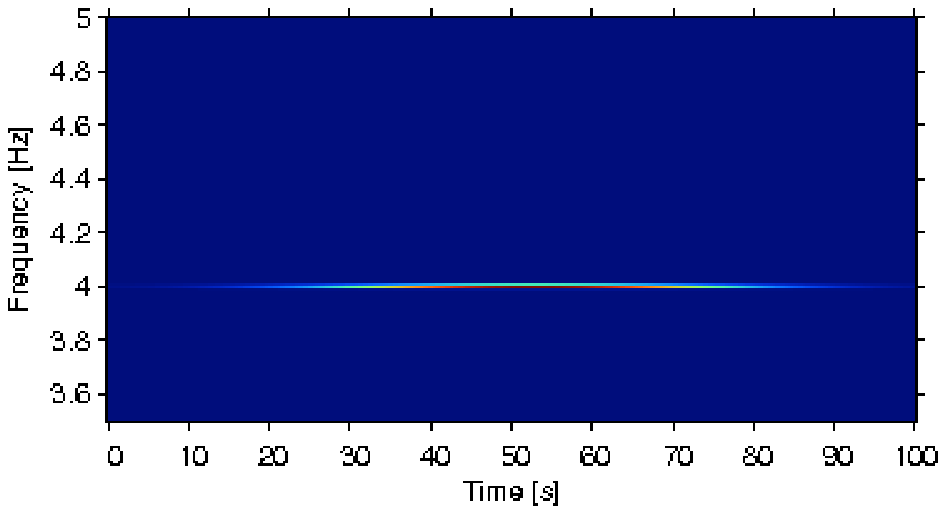}
\caption{Simulated QPO signal (a; upper panel) which is fully coherent and
single frequency but with random amplitude modulation. The middle panel (b)
shows the wavelet decomposition. This is not dissimilar to that from
the real data in Fig.~\ref{fig:zoom}b, though there are fewer large gaps and
the derived frequency is more stable. The bottom panel (c) shows the MP
decomposition of the same signal. This is clearly different to that of the real
data in Fig.~\ref{fig:zoom}c, ruling out this QPO shape. Colour coding of
a wavelet and MP map assumed the same as described in a caption of
Fig.~\ref{fig:wamp}.
}
\label{fig:sim1}
\end{figure}

\begin{figure}
\includegraphics[angle=0,width=0.475\textwidth]{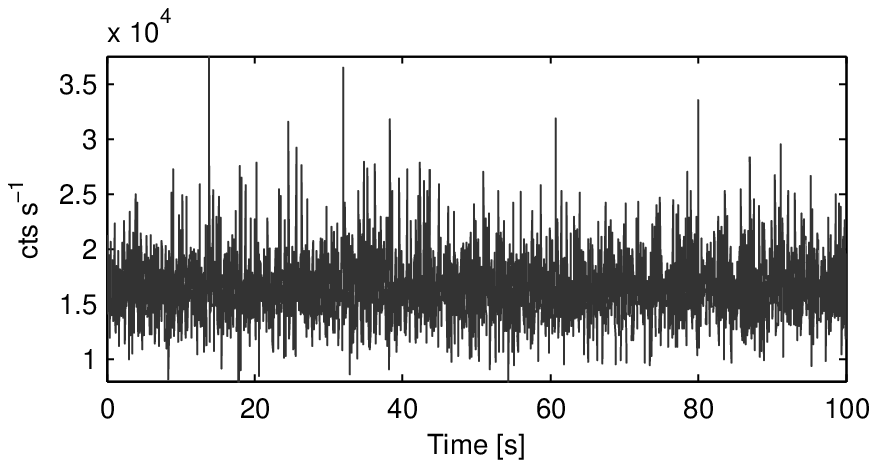}
\includegraphics[angle=0,width=0.475\textwidth]{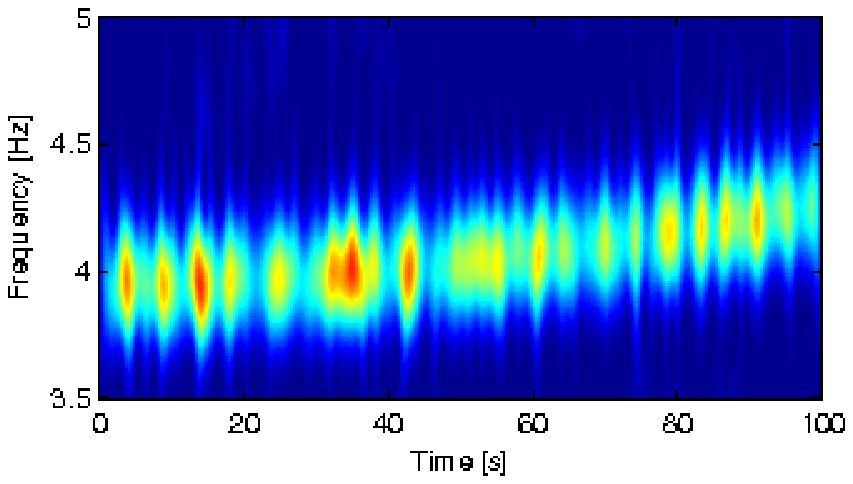}
\includegraphics[angle=0,width=0.475\textwidth]{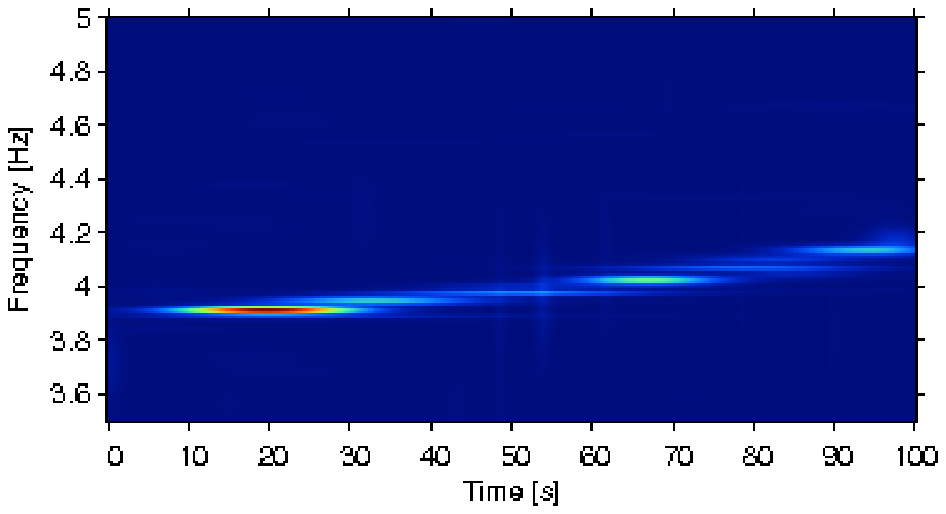}
\caption{As in Fig.~\ref{fig:sim1}, but for a simulated QPO signal
which is a fully coherent {\it chirp} signal, with frequency increasing with
time from 3.9 to 4.2~Hz. A smooth frequency drift is readily apparent in both
wavelet and MP maps, unlike the data shown in Fig.~\ref{fig:zoom}b and
\ref{fig:zoom}c. Colour coding of a wavelet and MP map assumed the same as
described in a caption of Fig.~\ref{fig:wamp}
}
\label{fig:chirp}
\end{figure}

\subsection{Coherent, single frequency signal with amplitude
  modulation}
\label{ss:coh}

Continuous amplitude modulation of a single frequency sinusoid will
result in a broad peak in a power spectrum. We simulate this by
generating a single sinusoid modulation $y(t)=A\sin(2\pi f_0 t)+C$
over 100~s with sampling time of $\Delta t = 2^{-5}$~s, $f_0=4$~Hz, and a
constant value of $C=1.65\times 10^4$ cts~s$^{-1}$. The amplitude $A$
is taken as a random variable from the lognormal distribution as fitted to the
XTE~J1550$-$564's histogram (see Fig.~\ref{fig:data}b and
Sect.~\ref{s:selection} for details). For $y(t)$ we perform
its exponential transformation, namely $\exp[y(t)]$, which accounts for a
non-linear type of variability present in X-ray light curves of accreting
black-hole systems (Uttley \& McHardy 2001; Uttley, McHardy \& Vaughan 2005). In
this point we ensured the signal would have a zero mean and its variance to be
such to obtain (after the above transformation and Poisson noise inclusion) a
similar degree of r.m.s. as observed for the entire XTE~J1550--564 data sample
as well as to match r.m.s. variability in the QPO band (3.5$-$5~Hz). 

Fig.~\ref{fig:sim1}a displays a simulated signal, whereas Fig.~\ref{fig:sim1}b
reveals the wavelet power spectrum. The spectrum is similar to
the XTE~J1550--564 wavelet map, but the MP decomposition is clearly different to
that in Fig.~\ref{fig:zoom}c. The first iteration returned a $\sim$85~s long
Gabor atom exactly at a frequency equal to the assumed $f_0$, carrying about
85\% of the signal energy and dominating over the rest of the fitted atoms.

Therefore the MP decomposition clearly rules out a QPO signal formed
solely from amplitude modulation as the real data cleanly show
multiple strong Gabor atoms with frequencies between 3.7--4.7~Hz.

\subsection{Coherent chirp signal with amplitude modulation}
\label{ss:chirp}

Physical models for the QPO involving blobs spiralling inwards should
produce a monotonically increasing frequency signal, otherwise known
as a chirp (e.g. Turner et al. 2006; Mhlahlo et al. 2007).

We repeate the experiment above, but with a chirp signal
where $y(t)=A\sin[2\pi f(t) t]+C$ and $f$ changes continuously
between 3.9 and 4.2 Hz. Fig.~\ref{fig:chirp} shows that again the amplitude
modulation can be easily captured in the wavelet map, but it leaves no impact on
frequency. In MP map, the continuously changing frequency is fitted
with separate atoms (see Sect.~\ref{ss:nomp} for more details on
Fig.~\ref{fig:chirp}c construction) which trace the drift. The lifetimes of the
individual atoms are around 10--20~s and show that the data can distinguish
between frequencies separated by $\sim 0.03-0.06$~Hz, i.e. approximately an
order of magnitude better than the wavelet technique at these frequencies.

Again, this MP map is quite unlike that of the real data shown in
Fig.~\ref{fig:zoom}c.

\begin{figure}
\includegraphics[angle=0,width=0.475\textwidth]{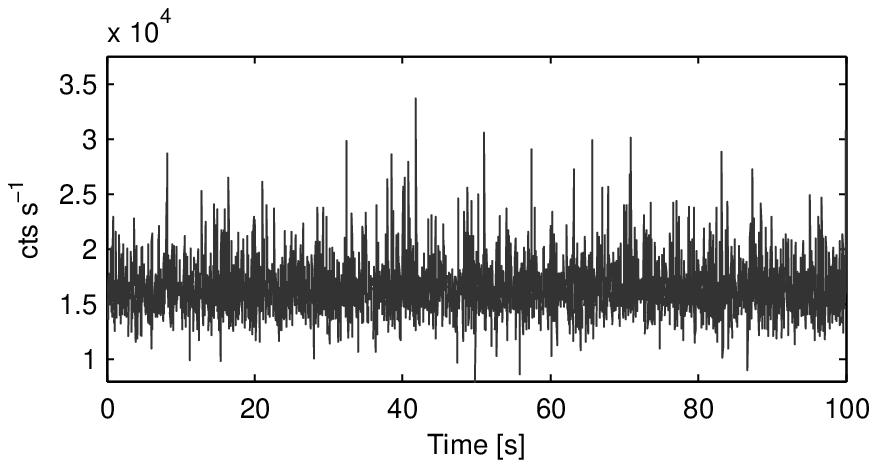} 
\includegraphics[angle=0,width=0.475\textwidth]{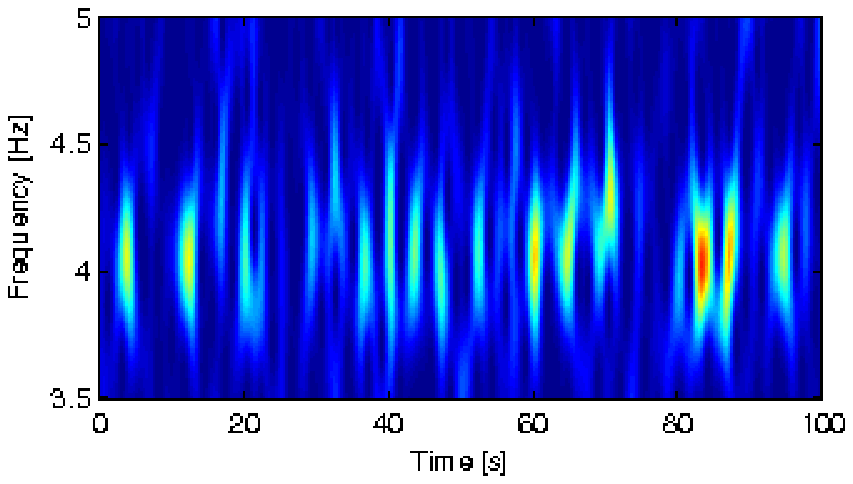}
\includegraphics[angle=0,width=0.475\textwidth]{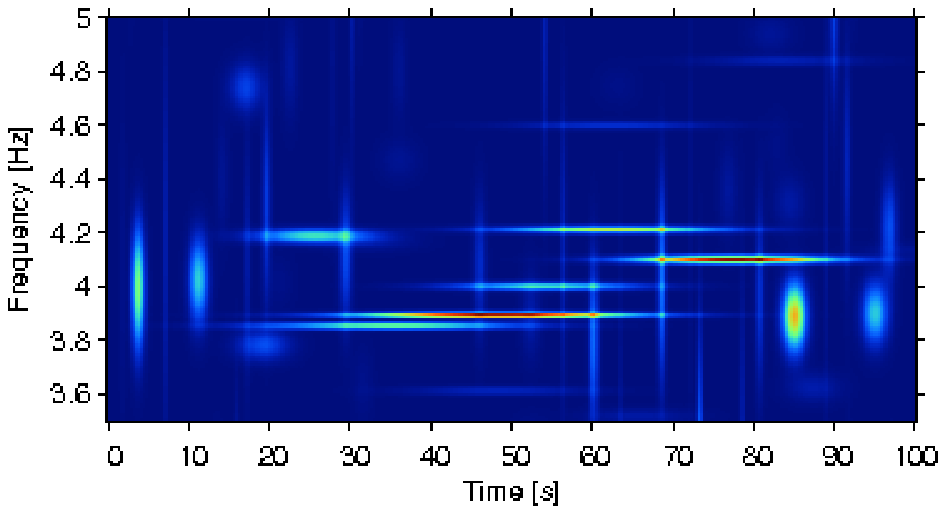}
\caption{As in Fig.~\ref{fig:sim1}, but for a simulated QPO signal
which is
composed of separate oscillations which last for a timescale randomly
distributed between 0-3~s followed by a wait time which is randomly
distributed between 0-2~s. Each oscillation has random phase and
amplitude but is at a fixed frequency of 4~Hz. The upper panel (a) shows
an exemplary light curve. The middle section (b) displays a corresponding
wavelet map, while the lowest panel (c) shows MP decomposition. Both these
are rather similar to those from the real data in Fig.~\ref{fig:zoom}b and
\ref{fig:zoom}c. Colour coding of
a wavelet and MP map assumed the same as described in a caption of
Fig.~\ref{fig:wamp}
}
\label{fig:random}
\end{figure}

\subsection{Short lifetime, random amplitude signals}
\label{ss:qpos}

The tests described above clearly show that the QPO signal in the data
is not given by a single frequency. There are multiple frequencies
present in the dataset, but these frequencies do not show any
systematic, long timescale trend such as a chirp.

This seems to fit most easily into a picture where the QPO is composed
of separate events with shape rather similar to that of the Gabor atom
i.e. a single frequency lasting for some timescale $t_{\rm qpo}$ which
varies. This would fit quite well into models where the QPO can be
randomly excited by turbulence, and last for some timescale before
getting damped out. The specific LF QPO model of Ingram et al. (2009) has a
vertical, Lense-Thirring precession of the hot
inner flow. This will be damped on a viscous timescale, which is of an
order $\sim 3$~s for typical parameters. 

\begin{figure}
\includegraphics[angle=0,width=0.475\textwidth]{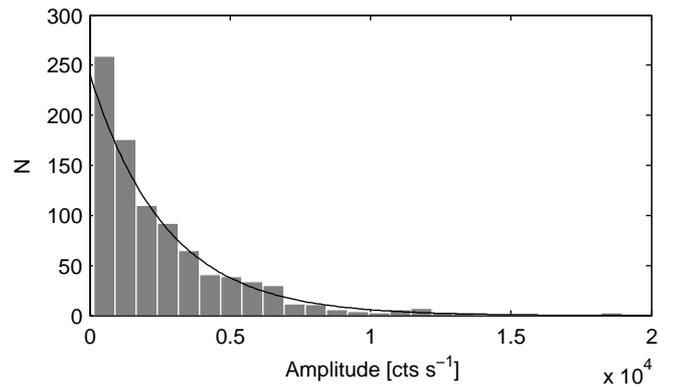}
\caption{The distribution of amplitudes for QPO Gabor atoms extracted from
Matching Pursuit analysis of XTE~J1550$-$564 light curve in 3.7$-$4.7~Hz
frequency band. The solid line denotes the exponential distribution fitted to
the data.}
\label{fig:h3}
\end{figure}

\begin{figure}
\includegraphics[angle=0,width=0.475\textwidth]{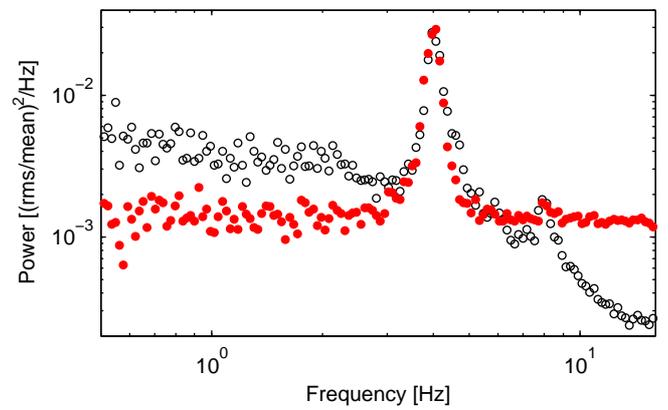}
\caption{The comparison between power spectral density caclulated for
XTE~J1550$-$564 (open black circles) and simulated signal as described in
Sect.~\ref{ss:qpos} (filled red circles).}
\label{fig:rmsqpo}
\end{figure}

Here, we approximate this physical picture by assuming that $t_{\rm qpo}$
varies uniformly between $(0,3]$~s and that after each QPO event there is a
waiting time of $t_{\rm break}$ randomly distributed in the $(0,2]$~s interval. 
Therefore the QPO is $\mbox{qpo}(t)=A\sin(2\pi f_0 t + \phi) + C$ where
$f_0=4$~Hz and a phase is randomised for every new QPO. As previously, we
assume the amplitude to be a random variable, however this time drawn from the
exponential
distribution,
\begin{equation}
 f(x; \mu) = \frac{1}{\mu} \exp\left(-\frac{x}{\mu}\right) ,
\end{equation}
fitted to the histogram of amplitudes derived for QPO atoms (3.7$-$4.7~Hz) from
the XTE~J1550$-$564 data, where we obtain $\mu=2692^{+245}_{-217}$ at the 99\%
confidence level (Fig.~\ref{fig:h3}). During the breaks we approximate signal
value by a random variable drawn from lognormal distribution and we follow the
signal transformation as described in Sect.~\ref{ss:coh}. 

The above construction of simulated signal allows us to account for a general
countrate distribution to resemble that in XTE~1550$-$564 light curve as well as
to approximate more adequately the amplitudes of simulated quasi-periodic
events. Fig.~\ref{fig:random}a displays an examplary 100~s signal obtained in
the above manner. During the simulation we assumed the r.m.s. variability in
3.5$-$5 Hz band to match closely 12\% as derived from the XTE~J1550$-$564 data.
In Fig.~\ref{fig:rmsqpo} we compare both Fourier power spectra, i.e. for the
X-ray source and simulated signal (here taken to be 1000~s long). The plots
indicate on very good qualitative agreement between the original and simulated
QPO feature in the frequency domain, though it is straighforward to note that
our simulation does not correctly reproduce broad-band red-noise variability
as observed in XTE~J1550$-$564.

Encouraged by the above agreement, we zoom in on the internal structure
of our simulation by performing wavelet and Matching Pursuit analysis as
previously (Fig.~\ref{fig:random}), and we find that both maps provide the
results which look very similar to that from the real data (cf.
Fig.~\ref{fig:zoom}b and c).

\begin{figure}
\includegraphics[angle=0,width=0.475\textwidth]{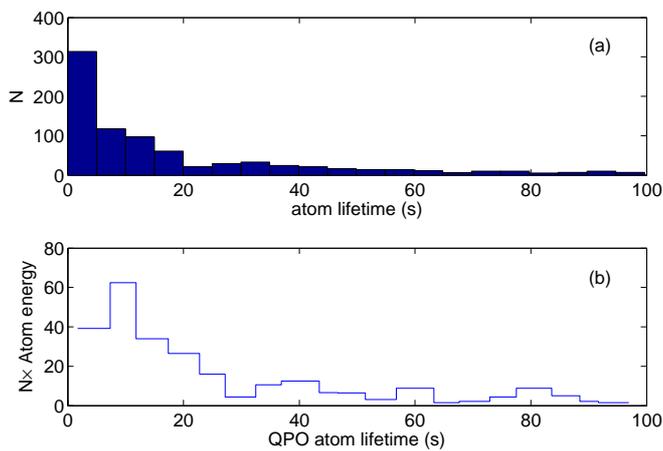}
\caption{As in Fig.~\ref{fig:hist_data}, histograms of atom lifetimes and
energies  for the simulated signal as described in Sect.~\ref{ss:qpos}. Bottom
panel refers to QPO atoms defined in 3.7$-$4.7 Hz band, excluding these events
of duration less than 0.5~s and of atom energy lower than in 2$-$3.7~Hz and
4.7$-$5~Hz bands.
}
\label{fig:hist_sim}
\end{figure}

More quantitatively, Fig.~\ref{fig:hist_sim} shows that the distributions of
atom lifetimes and energies are also remarkably well matched by the simulation.
Even though the signal is made from separate oscillations with life-time
of less than 3~s, an MP decomposition displays a number of longer lasting atoms,
e.g. of lifetimes between 40$-$80~s. Since this issue requires a better
understanding, we examined the MP response to the set of various test signals
containing different noise levels, amplitude distribution, variable frequency
and/or constant phase for separate qpo$(t)$ events. In conclusion, we have
noticed that the longer lasting atoms can be fitted to the data by an MP
algorithm making use of an {\it atom concatenation} process. In this case, a
number of shorter lasting oscillations characterized by very similar phase and
frequency (not amplitude!) can be linked together into one longer atom.
Therefore, e.g. a 60~s long atom can describe, say, three to five $\sim$3~s
oscillations, and our inability to distinguish among them refers to the
limited time-frequency resolution of the Matching Pursuit algorithm.

Similarly, while the simulation has only one constant frequency signal, there is
a spread in frequency in the MP decomposition which matches that seen in the
data.

The low-frequency QPO seen in the real data can be well described by a
signal which consists of random, discrete events at a single frequency, with a
spread in lifetime and amplitude.

\section{Discussion and Conclusion}
\label{s:discuss}

Quasi-periodic oscillations observed in the power spectra of many accreting
compact objects still remain the most intriguing puzzle. Since
high-frequency QPOs are generally thought as the most attractive landmark
feature used in testing the behaviour of matter in the very strong gravitational
field (e.g. Kluzniak \& Abramowicz 2001; Rebusco \& Abramowicz 2006), on the
contrary, the low-frequency QPOs have been found to be more enigmatic in their
nature (Casella et al. 2006). For the latter, a quantitative analysis has been a
dominant tool towards our comprehension of the global QPO properties (e.g.
Cui et al. 1999; Sobczak et al. 2000).

The presence of broad peaks in the Fourier power spectra fitted with
Lorentzian profiles suggests a possible mechanism of X-ray engine at work:
damping of locally emitted radiation at the very short time-scales (e.g. Misra
\& Zdziarski 2008). The position of QPO peak frequency varies in time, generally
being drifting between 0.1$-$10~Hz (van der Klis 2005, 2006). It has been
suggested it may be correlated with both underlying geometry of the inner flow,
and accretion rate, thus with the spectral state of the source (e.g. Di Matteo 
\& Psaltis 1999; Dutta et al. 2008; Qu et al. 2010). The r.m.s. variability of
LF QPOs is also significant and in some cases exceeds 10\%.

The successful modelling of this QPO phenomena requires a sophisticated
approach which would include acceptable model's results not only in the
reproductions of the light curve and Fourier PSD, but also in providing
an agreement in a function of energy (e.g. Sobolewska {\.Z}ycki 2006). Such
attempts have been so far conducted on many occassions (e.g. Giannios \& Spruit
2004; Machida et al. 2006; Machida \& Matsumoto 2008). Fundamental and dynamical
Fourier analysis of X-ray light curves gained a leading position in spoting the
existence and evolution of QPOs (e.g. Galleani et al. 2001; Pottschmidt et al.
2003; Barret et al. 2005) but failed in delivering the satisfactory answer on
the physical structure due to insufficient resolution. On the hand,
the application of new time-frequency methods for X-ray time-series analysis
(e.g. by wavelet approach) initiates a novel look at the data and brings an
opportunity for testing QPO models in a new space of parameters (e.g.
Steiman-Cameron et al. 1997; Parker 1998; Lachowicz \& Czerny 2005).

In this paper we have applied the Matching Pursuit method for signal
decomposition and shown how MP technique can give much better resolution in
the time-frequency
plane to study the evolution of the low-frequency QPO seen from black hole
binary systems. We have used data from the bright transient systems XTE
J1550$-$564 to
illustrate this with an observation taken in the very high state (aka the steep
power law state) where the QPO is strong and relatively
coherent.

The Mathching Pursuit decomposition clearly rules out several potential origins
for the broadening of the QPO signal. The `quasi' nature cannot be due to a
single frequency, coherent oscillation with amplitude modulation, nor can it be
a long timescale ($\sim$100~s) systematic change in frequency (chirp signal).
Instead, it is well matched by a series of discrete oscillations, with
life-times between 0--3~s, each with random phase and amplitude but fixed
frequency.

Physically, this is consistent with models of the QPO where it is excited by
random turbulence.  In particular, the Lense-Thirring precession model of Ingram
et al. (2009) can fit nicely into this picture, where the vertical precession is
excited by the large scale stochastic turbulence generated in the accretion
flow. The timescale for decay of the vertical precession should be related to
the viscous timescale of the flow. In the Ingram et al. (2009) model, a 4~Hz QPO
would be associated with an outer radius of the hot flow of $\sim10R_g$, so
should have a viscous timescale of $\sim 2-3$~s for the typical parameters used
by Ingram et al. (2009).

The application of Matching Pursuit opens a brand new opportunity for
studies of LF QPOs in the time-frequency domain. It encourages for revision of
archival data of other BHB which display narrow QPOs in their Fourier power
spectra. By increased time-frequency resolution provided by MP it is possible
to diferentiate amongst various QPO models, as demonstrated in this paper.
Therefore, the Matching Pursuit analysis may bring us much closer towards
understanding of the underlying physical properties of the QPO phenomenon.

\begin{acknowledgements}
We would like to thank the anonymous referee for useful remarks that helped
to improve our paper; to Say Song Goh, Piotr Durka, Jarek Zygierewicz, Sylvie
Roques for many friendly discussions on Matching Pursuit algorithm; to Adam
Ingram, Didier Barret, Phil Uttley for comments; Marek Michalewicz and Kevin
Jeans for reading the manuscript and helpful remarks; and to Izabela
Dyjeci\'{n}ska for graphic advice. The research has made use of data obtained
through the High Energy Astrophysics Science Archive Research Center Online
Service, provided by the NASA/Goddard Space Flight Centre. 
\end{acknowledgements}

\appendix

\section{Time-Frequency Analysis}
\label{s:tf}

\subsection{Spectrogram and wavelet analysis}

In X-ray time analysis the wavelet transformation constitutes an alternative
tool for signal decomposition when confronted with the classical approach
based on the Fourier transform (PSD). In the latter, a signal is analysed with
the basis functions $\psi \propto e^{{\rm i}ft}$. It detects
strictly periodic modulations (with period of $1/f$) buried in the signal.
Therefore, no information about time-frequency properties is available.

A simple solution to this problem has been found via the application of
Short-Term Fourier Transform (STFT). Basic concept standing for this method is
to divide signal into small segments and calculate for each of them Fourier
transform. The signal analysis performed using {\it spectrogram}, i.e.
$|$STFT$|^2$ often blurs all signal oscillations of life-times
less than the duration of a segment causing a loss of significant information.

The wavelet analysis treats this problem more accurately as a basic concept
standing for its usefulness relay on probing time-frequency plane at
different frequencies with different resolutions. The wavelet power spectrum
for a discrete signal can be defined as the normalised square of the modulus of
the wavelet transform $\xi |w_n(a_m)|^2$ where $\xi$ stands for
normalisation factor and $w_n(a_m)$ a discrete form of the continuous wavelet
transform (Farge 1992; Torrence \& Compo 1998; Lachowicz \& Czerny 2005). It is
attractive to adopt a Morlet wavelet to probe quasi-periodic modulations.
Since Morlet wavelet oscillates due to a term $\propto e^{{\rm i}t}$, it is
perfectly suited for this task.

\subsection{Matching Pursuit algorithm}
\label{ss:mp}

\subsubsection{Parlez-vous fran\c{c}ais?}

Signal time-frequency analysis can be compared to speaking in a foreign
language. In each language we use words. Words are needed to express
our thoughts, problems, ideas, etc. By a smart selection of proper words we
can say and explain whatever we wish. A whole collection of words can be
gathered in the form of a dictionary. One can express simple thoughts
using a very limited set of words from a huge dictionary (a subset). The same
can be applied to a time-series analysis. In order to describe the signal one
needs to use a minimum available set of functions -- orthonormal basis
functions.

Description of the signal $x_n$ which uses only basis functions $g_n$ 
$(n=1,...,N')$ is therefore limited. This is often the case with wavelet
transform. In order to improve signal description, one can increase the size of
a basic dictionary by adding extra functions. Such a dictionary's enlargement
introduces a {\em redundancy} what is a natural property of all, for instance,
foreign languages. The most reliable signal description can be achieved by a
signal
approximation:
\begin{equation}
 x \approx \sum_{n=1}^{M<N'} a_{\gamma,n} g_{\gamma,n}
\end{equation}
where coefficients $a_{\gamma,n}$ are defined simply as the inner products of
basis functions $g_{\gamma,n}$ with a signal:
\begin{equation}
 \langle x, g_{\gamma,n} \rangle = \int_{-\infty}^\infty x(t) g_{\gamma}(t) dt
 \approx \sum_{n=1}^{M<N'} x_n g_{\gamma,n} 
\end{equation}
and $\gamma$ refers to a set of parameters of function $g$. 

In each approximation, a relatively small number of functions is generally
required.  For each case the signal is projected onto a set of basis functions,
therefore, in other words, its representation is always fixed. However, what if
one may approach the signal decomposition problem from the other side, i.e. to
fit the singal representation to the signal itself? One can achieve it by
choosing from a huge dictionary of pre-defined functions a subset of functions
which would best represent the signal. Such solution had been proposed by
Mallat \& Zhang (1993) and is known under the name of {\it Matching Pursuit}
algorithm.

\subsubsection{Algorithm}
\label{ssa:algo}

Matching Pursuit (MP) is an iterative procedure which allows to choose from a
given, redundant dictionary $D$, where $D=\lbrace g_1, g_2,...,g_n\rbrace$
such that $||g_i||\!=\!1$, a set of $m$ functions 
$\lbrace g_i\rbrace_{m<n}$ which match the signal as well as possible. $D$
is defined as a family (not a basis) of time-frequency waveforms that can be
obtained by time-shifting, scaling and, what is new comparing to wavelet
transform, modulating of a single even function $g\in L^2(\Re)$; 
$\Vert g \Vert=1$;
\be
   g_\gamma = \frac{1}{\sqrt{a}} g\left(\frac{t-b}{a}\right) e^{i f t} 
\ee
where an index $\gamma$ refers to a set of parameters:
\be
    \gamma=\lbrace a,b,f \rbrace .
\label{mupar}
\ee
Each element of the dictionary is called an
{\it atom} and usually is defined by
a Gabor function, i.e. a Gaussian function modulated by a cosine of
frequency $f$:
\begin{equation}
 g_\gamma(t)\!=\!K_\gamma(t) G_\gamma(t)\! =\!
 K_\gamma(t)
 	     \exp\left[-\pi\left(\frac{t-b}{a}\right)^2\right]
 	     \cos\left( 2\pi ft + \phi \right)
\label{gabor}
\end{equation}
where
\begin{equation}
 K_\gamma(t) = \left( {\sum_{t=-\infty}^\infty G_\gamma(t)^2} \right)^{-1/2} 
\end{equation}
and $a$ denotes the atom's scale (duration; in this paper also referred to as
{\it
atom's life-time}) and $b$ -- time position. A selection of Gabor functions for
purposes of MP method is fairly done as they provide optimal joint
time-frequency localisation (e.g. Mallat \& Zhang 1993; Cohen 1995; Flandrin
1999).

In the first step of MP algorithm, a function $g_{\gamma,1}$ is chosen from $D$
which matches the signal $x$ best. In practice it simply means that a
whole dictionary $D$ is scanned for such $g_{\gamma,1}$ for which a scalar
product $|\langle x,g_{\gamma,1}\rangle|\in\Re$ is as large as possible.
Therefore, a signal can be decomposed into:
\begin{equation}
    x = \langle x,g_{\gamma,1}\rangle g_{\gamma,1} + R^1x
\end{equation}
where a $R^1x$, a residual signal, has been denoted. In the next step,
instead of $x$, the remaining signal $R^1x$ is decomposed by finding a
new function, $g_{\gamma,2}$, which matches $R^1x$ best. Such consecutive
steps can be repeated $p$ times where $p$ is a given, maximum number of
iterations for signal decomposition. In short, MP procedure can be denoted as:
\begin{equation}
      \left\{ \begin{array}{l}
		R^0x = x\\
		R^ix = \langle R^ix,g_{\gamma,i}\rangle g_{\gamma,i} +
				R^{i+1}x \\
		g_{\gamma,i} = \arg\max_{g_{\gamma,i'}\in D} 
				|\langle R^ix,g_{\gamma,i'}\rangle| .
	      \end{array}
      \right.
\label{induc}
\end{equation}
Davis et al. (1994) showed that for $x\in H$ the residual term $R^ix$
defined by the induction equation (\ref{induc}) satisfies:
\begin{equation}
	\lim_{i\rightarrow +\infty} \Vert R^ix\Vert = 0
\end{equation}
i.e., when $H$ is of finite dimension, $\Vert R^ix\Vert$ decays to zero.
Hence, for a complete dictionary $D$, the Matching Pursuit procedure converges
to
$x$:
\begin{equation}
	x = \sum_{i=0}^\infty \langle R^ix,g_{\gamma,i}\rangle g_{\gamma,i}
\label{mpxt}
\end{equation}
where orthogonality of $R^{i+1}x$ and $g_{\gamma,i}$ in each step implies
energy conservation:
\begin{equation}
  \begin{array}{ll}
  \Vert x \Vert^2\! & = 
                           \sum_{i=0}^{p-1}
                           |\langle R^ix,g_{\gamma,i}\rangle|^2
                           + \Vert R^px\Vert^2 \\
                       & = \sum_{i=0}^{\infty} 
                           |\langle 
                           R^ix,g_{\gamma,i}\rangle|^2 \ .
    \end{array}
\label{engcon}
\end{equation}

\subsubsection{Time-frequency representation}

From (\ref{mpxt}) one can derive a time-frequency distribution of signal's
energy by adding a Wigner-Ville distribution of selected atoms $g_\gamma$:
\begin{equation}
  \begin{array}{ll}
  (Wx)(t,f) & = \sum_{i=0}^\infty |\langle R^ix,g_{\gamma,i}\rangle|^2 
                  (Wg_{\gamma,i})(t,f) + \\
              &   \sum_{i=0}^\infty \sum_{j=0,j\ne i}^\infty
                  \langle R^ix,g_{\gamma,i}\rangle
                  \langle R^jx,g_{\gamma,j}\rangle^\ast \ \ \times \\ \nonumber
 	      &   \ \ \ \ \ \ \ \ \ \ \ \ \ \ \ \ \ \ \ \ \ 
                  \ \ \ \ \ \ \ \ \ \ \ \ \ \ \ \times \ \ 	
	          (W[g_{\gamma,i},g_{\gamma,j}])(t,f)
\end{array}
\end{equation}
The double sum present in the above
equation corresponds to cross-terms present in the Wigner-Ville distribution.
In MP approach, one allows to remove these terms
directly in order to obtain a clear picture of signal energy distribution in
the time-frequency plane $(t,f)$. Thanks to that, {\it energy density} of
$x$ can be defined as follows:
\begin{equation}
   (Ex)(t,f) = \sum_{i=0}^\infty
                 |\langle R^ix,g_{\gamma,i}\rangle|^2
		 (Wg_{\gamma,i})(t,f) .
\label{mpeng}
\end{equation}
Since Wigner-Ville distribution of a single time-frequency atom $g_\gamma$
satisfies:
\begin{equation}
   \int_{-\infty}^{\infty} \int_{-\infty}^{\infty}
   (Wg_{\gamma,i})(t,f) dt df = \Vert g_\gamma \Vert^2 = 1
\label{wvone}
\end{equation}
thus combining (\ref{wvone}) with energy conservation of the MP expansion
(\ref{mpxt}) yields:
\begin{equation}
   \int_{-\infty}^{\infty} \int_{-\infty}^{\infty}
   (Ex)(t,f) dt df = \Vert x \Vert^2
\end{equation}
what justifies interpretation of $(Ex)(t,f)$ as the energy density of
signal $x$ in the time-frequency plane.

In practical computation of MP algorithm, sampling of a discrete signal
of length $N=2^L$ is governed by an additional parameter $j$:
an octave (Mallat \& Zhang 1993). The scale $a$ corresponding to an atom's
scale is derived from a dyadic sequence of $a=2^j, 0\le j\le L$. Thus,
parameters $b$ and $f$, corresponding to atom's position in time and frequency,
respectively. By {\it resolution} in the MP one may understand the
distance between centres of atoms neighboring in time or in frequency. A dyadic
sampling pattern proposed by Mallat \& Zhang has been found to suffer from a
statistical bias introduced in the atom parametrisation (\ref{mupar}). Durka et
al. (2001) proposed a solution to this problem by application of {\it stochastic
dictionaries}. In this approach the parameters of dictionary's waveforms are
randomised before each decomposition by drawing their values from continuous
ranges.

In this paper in all time-series analyses with MP method, we decompose
signal in 500 iterations using, if not otherwise stated, stochastic
dictionaries composed of $3\times 10^6$ atoms. By the term {\em
strongest atoms} we will refer to these Gabor atoms for which the
calculated inner product with a signal has been largest (in practice
it refers to the atoms fitted within first 1--10 iterations).

\subsubsection{Number of MP decompositions for a single signal}
\label{ss:nomp}

Durka et al. (2001) pointed at a significant aspect of Matching Pursuit method:
an ability to represent signals with continuously changing frequency. In the
standard approach  one aims at calculation of a {\em single} MP decomposition
for
a given signal using a large redundant dictionary of pre-defined functions
(e.g. $3\times 10^6$ atoms). When the signal contains continuously
changing frequency in time a single MP decomposition will provide energy density
plot to be composed of a number of single Gabor atoms. Therefore, such a view
remains far different from expectations (see exemplary illustration in Durka
et al. 2001, 2007).

On the way of experiments Durka et al. (2001) discovered that by introduction of
stochastic dictionaries and averaging a larger number,
say 100, of the MP energy density plots together one can uncover
continuously changing signal frequency. The trick is to use for every MP
decomposition performed for the same signal a new stochastic dictionary and of
much smaller size, e.g. composed of $5\times 10^4$ atoms only. It has been
applied by us in order to create Fig.~\ref{fig:chirp}c.

In this paper we have checked that for all our data segments the MP results
are insensitive to the way of MP computation, i.e. irrespective of the
dictionary size used and the application of aforementioned MP map averaging
process. In this way, we are able to exclude a scenario in which QPOs are
being generated due to long timescale oscillations of continuously changing
frequency (see Section~\ref{ss:chirp} for details).

\end{document}